\documentclass[11pt,twoside]{article} \usepackage[unicode]{hyperref}
\usepackage[margin=1in,centering]{geometry}\newcommand\aut {Leonid A.~Levin}
\newcommand\ttl {Enumerable Distributions, Randomness, Dependence}
\usepackage {amsmath,amssymb,amsthm,bm,mmap} \usepackage [T1,T2A] {fontenc}
\usepackage[utf8]{inputenc}\usepackage[russian,english]{babel}
\usepackage{microtype}
\begin{document}

 \newcommand\p[1]{\vspace{-1ex}\paragraph{#1}\hspace{-1em}}
 \newtheorem{thm}{Theorem} \newtheorem{lem}{Lemma} \newtheorem{cor}{Corollary}
 \newtheorem{prp}{Proposition} \newtheorem{dfn} {Definition}
 \newtheorem{rmr} {Remark}\newcommand\BR{\begin{rmr}}\newcommand\ER {\end{rmr}}
 \newcommand\BD {\begin{dfn}} \newcommand\ED {\end{dfn}}
 \newcommand\BL {\begin{lem}} \newcommand\EL {\end{lem}}
 \newcommand\BT {\begin{thm}} \newcommand\ET {\end{thm}}
 \newcommand\BP {\begin{prp}} \newcommand\EP {\end{prp}}
 \newcommand\BC {\begin{cor}} \newcommand\EC {\end{cor}}
 \newcommand\BPR {\begin{proof}} \newcommand\EPR {\end{proof}}
 \newcommand\BE {\begin{enumerate}} \newcommand\EE {\end{enumerate}}

 \newcommand\hreff[1] {{\footnotesize\href{https://#1}{https://#1}}}
 \newcommand\emm[1]{{\ensuremath{#1}}} \newcommand\trm[1]{{\bf\em #1}}
 \newcommand\emb[1]{{\ensuremath{\mathbf{#1}}}}\frenchspacing

 \newcommand\ov[1]{{\overline{#1}}} \newcommand\un[1]{{\underline{#1}}}
 \newcommand\floor[1]{{\lfloor#1\rfloor}} \newcommand\ceil[1]{{\lceil#1\rceil}}
 \newcommand\edf{{\raisebox{-3pt}{$\,\stackrel{\text{\tiny df}}{=}\,$}}}
 \newcommand\tld[1]{{\raisebox{-1pt}{$\widetilde{#1}$}}}
 \newcommand\wht[1]{{\raisebox{-1pt}{$\widehat{#1}$}}}
 \newcommand\lea{\prec}\newcommand\gea{\succ} \newcommand\eqa{\asymp}
 \newcommand\lel{\lesssim}\newcommand\gel{\gtrsim}\newcommand\eql{\sim}

 \renewcommand\i {{\emb i}} \renewcommand\d   {{\emb d}}
 \newcommand\M  {{\emb M}}    \newcommand\tb {{\emb t}} 
 \newcommand\St {{\emb S}}    \newcommand\I   {{\emb I}}
 \newcommand\T  {{\emb T}}    \newcommand\m   {{\emb m}}
 \newcommand\KM {{\emb{KM}}}  \newcommand\one {{\emb 1}}
 \newcommand\K  {{\emb K}}    \newcommand\Ki  {{\wht\K}}
 \newcommand\mf {{\wht\m}}    \newcommand\If {{\wht\i}}
 \newcommand\N {{\emm{\mathbb N}}} \newcommand\Q  {{\emm{\mathbb Q}}}
 \newcommand\R {{\emm{\mathbb R}}} \newcommand\Ks {{\raisebox{2pt}{\emm\chi}}}
 \newcommand\Es{{\emm{\bf\cal E}}} \newcommand\Ess {{\tld\Es}}

 \renewcommand\a {{\emm\alpha}}  \renewcommand\b {{\emm\beta}}
 \renewcommand\l {{\emm\lambda}} \renewcommand\r {{\emm\rho}}
 \newcommand\g   {{\emm\gamma}}  \newcommand\dl  { {\emm\delta}}
 \newcommand\w   {{\emm\omega}}  \newcommand\W   {{\emm\Omega}}
 \newcommand\ph  {{\emm\varphi}} \newcommand\ie {{\em i.e., }}
 \newcommand\eg  {{\em e.g., }}  \newcommand\re {{\em r.e. }}

 \title {\vspace*{-6pc}\ttl} \date{} \author{\aut\\
 Boston University\thanks {Computer Science dept., 111 Cummington Mall,
 Boston, MA 02215; Home page: \hreff{www.cs.bu.edu/fac/Lnd}}}\maketitle

\vspace*{-3pc}\begin{flushright}\parbox{1pc}{\begin{tabbing}
           С этой безмерностью в мире мер.\\*\em -- Марина Цветаева
\footnotemark\end{tabbing}}\end{flushright}\footnotetext
          {Measureless in this world of measures. -- Marina Tsvetaeva}

\vspace*{-2pc}\begin{abstract}\noindent Mutual information $\I$ in infinite
sequences (and in their finite prefixes) is essential in theoretical analysis
of many situations. Yet its right definition has been elusive for a long time.
I address it by generalizing Kolmogorov Complexity theory from measures
to {\bf semimeasures} \ie infimums of sets of measures. Being concave
rather than linear functionals, semimeasures are quite delicate to handle.
Yet, they adequately grasp various theoretical and practical scenaria.

A simple lower bound $\i(\a:\b)\edf\sup_{x\in\N}(\K(x)- \K(x|\a)-\K(x|\b))$
of information turns out tight for Martin-L\"of random $\a,\b\in\{0,1\}^\N$.
For all sequences $\I(\a:\b)$ is characterized by the minimum of $\i(\a':\b')$
over random $\a',\b'$ with $U(\a'){=}\a$, $U(\b'){=}\b$.\end{abstract}

\section {Introduction}

Kolmogorov Information theory applies to individual objects, in contrast to
Shannon theories that apply to the models of processes that generated such
objects. It thus has a much wider domain since many objects (\eg Shakespeare
plays) have no realistic generation models. For completed objects, such as
integers, the concept is simple and robust: $\I(x:y)=\K(x)+\K(y)-\K(x,y)$.

Yet, the concept is also needed for emerging objects, such as, \eg prefixes of
infinite sequences. Encoding prefixes as integers distorts the information by
specifying their (arbitrary) cut-off point. This cut-off information is not a
part of the original sequence and can be smaller in a longer prefix. In fact,
this distortion can overwhelm the actual mutual information between the
sequences.

This issue complicates many studies forcing one to use (as, \eg in \cite{fi})
concepts of information that are merely lower bounds, differ between
applications, and known not to be tight.

For the related concept of rarity (randomness deficiency) Per Martin-L\"of
proposed an extention that works well for infinite sequences under computable
distributions. Yet, computability of distributions requires a running time
limit for the processes generating them. Such limits then must be accounted for
in all formulas, obscuring the simplicity of purely informational values, at a
great cost to elegance and transparency. Without such limit many important
distributions are only lower-enumerable (r.e.). For instance, universal
probability {\M} is the largest within a constant factor \re distribution. It
is extraordinarily flat: all sequences are random with respect to it.

Yet {\M} is instrumental in defining other interesting distributions. In
particular, Mutual Information in two sequences is their \trm {dependence}, \ie
rarity with respect to the distribution $\M\otimes\M$ generating them
independently with universal probability each. R.e. distributions are of
necessity semimeasures: concave rather than linear functionals. Semimeasures
also are relevant in more mundane and widespread situations where the specific
probability distribution is not fully known (\eg due to interaction with a
party that cannot be modeled). They require much more delicate handling than
measures. This article considers many subtleties that arise in such
generalization of complexity theory. The concept of rarity for such
distribution considered here respects randomness conservation inequalities and
is the strongest (\ie largest) possible such definition. The definition of
mutual information arising from this concept is shown to allow rather simple
descriptions.

\section {Conventions and Background}

Let \R, \Q, \N, $\St{=}\{0,1\}^*$, $\W{=}\{0,1\}^\N$ be, respectively,
the sets of reals, rationals, integers, finite, and infinite binary sequences;
$x_{[n]}$ is the $n$-bit prefix and $\|x\|$ is the bit-length of $x{\in}\St$;
for $a{\in}\Re^+$, $\|a\|{\edf}|\,\ceil{\log a}{-}1|$. A function $f$
and its values are \trm {enumerable} or \trm\re ($-f$ is \trm{co-r.e.}) if its
subgraph $\{(x,t):t<f(x)\}$ is r.e., i.e. a union of an \re set of open balls.
$X^+$ means $X\cap\{x{\ge}0\}$. \trm {Elementary} ($f{\in}\Es$) are functions
$f:\W\to\Q$ depending on a finite number of digits; $\one\in\Es$ is their
unity: $\one(\a)=1$. $\tld E$~is the set of all supremums of subsets of $E$.
$f{\uparrow}$ for $f:\W\to\R$, denotes $\sup\{g:f>g\in\Es\}$.

\trm {Majorant} is an \re function largest, up to constant factors, among
\re functions in its class.\\ ${\lea}f$, ${\gea}f$, ${\eqa}f$, and ${\lel}f$,
${\gel} f$, ${\sim}f$ denote ${\le}f{+}O(1)$, ${\ge}f{-}O(1)$, ${=}f{\pm}O(1)$,
and ${\le}f{+}O(\|f{+}1\|)$,\\ ${\ge}f{-}O(\|f{+}1\|)$, ${=}f{\pm}O(\|f{+}1\|)$,
respectively. $[A]\edf1$ if statement $A$ holds, else $[A]\edf0$.

When unambiguous, I identify objects in clear correspondence: \eg prefixes with
their codes or their sets of extensions, sets with their characteristic
functions, etc.

\subsection {Integers: Complexity, Randomness, Rarity}

Let us define Kolmogorov \trm {complexity} $\K(x)$ as $\|\m(x)\|$ where
$\m:\N\to\R$ is the \trm {universal distribution}, \ie a majorant \re function
with $\sum_x\m(x){\le}1$. It was introduced in \cite{ZL}, and noted in \cite
{L73,L74,g74} to be a modification of the least length of binary programs for
$x$ defined in \cite {K65}. The modification restricts the domain $D$ of the
universal algorithm $u$ to be prefixless. While technically different, {\m}
relies on intuition similar to that of \cite {Sol}. The proof of the existence
of a majorant function was a direct modification of \cite {Sol, K65} proofs
which have been a keystone of the informational complexity theory.

For $x{\in}\N,y{\in}\N$ or $y{\in}\W$, similarly, $\m(\cdot|\cdot)$ is a
majorant \re real function with $\sum_x\m(x|y){\le}1$; $\K(x|y)\edf\|\m(x|y)\|$
($=$ the least length of prefixless programs transforming $y$ into $x$).

\cite {K65} considers \trm {rarity} $\d(x)\edf\|x\|{-}\K(x)$ of uniformly
distributed $x{\in}\{0,1\}^n$.\\ Our modified {\K} allows extending this to
other measures $\mu$ on~$\N$. A $\mu$-test is $f:\N\to\R$ with mean
$\mu(f){\le}1$ (and, thus, small values $f(x)$ on randomly chosen~$x$). For
computable $\mu$, a majorant \re test is $\tb(x)\edf\m(x)/\mu(x)$. This
suggests defining $\d_\mu(x)$ as $\|\ceil{\tb(x)}\|\eqa \|\mu(x)\|-\K(x)$.

\subsection {Integers: Information}

In particular, $x{=}(a,b)$ distributed with $\mu{=}\m\otimes\m$, is a pair
of two independent, but otherwise completely generic, finite objects.
Then, $\I(a:b)\edf\d_{\m\otimes\m}((a,b)){\eqa}\K(a){+}\K(b){-}\K (a,b)$ measures
their \trm {dependence} or \trm {mutual information}. It was shown
(see \cite{ZL}) by Kolmogorov and Levin to be close (within
${\pm}O(\log\K(a,b))$) to the expression $\K(a){-}\K(a|b)$ of \cite{K65}.
Unlike\\ this earlier expression (see \cite {g74}), our {\I} is symmetric and
monotone: $\I(a:b)\lea\I((a,x):b)$ (which will allow extending {\I} to $\W$);
it equals $\eqa\K(a)-\K(a|\ov b)$, where by $\ov b$ we will denote $(b,\K(b))$.
\\ (The $\I_z$ variation of $\I$ with all algorithms accessing oracle $z$,
works similarly.)\\ $\I$ satisfies the following Independence Conservation
Inequalities \cite{L74,L84}:\\ For any computable transformation $A$ and measure
$\mu$, and some family $t_{a,b}$ of $\mu$-tests\vspace{-5pt}
 \[(1)\ \I(A(a):b)\lea \I(a:b); \hspace{4pc} (2)\ \I((a,w):b)\lea
 \I(a:b)+\log t_{a,b}(w).\vspace{-5pt}\]
 (The $O(1)$ error terms reflect the constant complexities of $A,\mu$.)
 So, independence of $a$ from $b$ is preserved in random processes,
 in deterministic computations, their combinations, etc. These
inequalities are not obvious (and false for the original 1965 expression
$\I(a:b){=}\K(a){-}\K(a/b)$~) even with $A$, say, simply cutting off
half of $a$. An unexpected aspect of $\I$ is that $x$ contains all
information about $k{=}\K(x)$, $\I(x:k)\eqa\K(k)$, despite
$\K(k|x)$ being ${\sim}\|k\|$, or ${\sim}\log\|x\|$ in the worst case
\cite{g74}. One can view this as an "Occam Razor'' effect: with no
initial information about it, $x$ is as hard to obtain as its simplest
($k$-bit) description.

\subsection {Reals: Measures and Rarity}\label{ML}

\p {A measure} on $\W$ is a function $\mu(x){=}\mu(x0){+}\mu
(x1)$, for $x{\in}\St$. Its mean $\mu(f)$ is a functional on \Es, linear:
$\mu(cf{+}g){=}c\mu(f){+}\mu(g)$ and \trm {normal:} $\mu(\pm\one){=}\pm1$,
$\mu(|f|)\ge0$. It extends to other functions, as usual.
An example is $\l(x\W)\edf 2^{-\|x\|}$ (or $\l(x)$ for short).
 I use $\mu_{(\a)}(A)$ to treat the expression $A$ as a
 function of $\a$, taking other variables as parameters.

$\mu$-\trm{tests} are functions $f\in\Ess$, $\mu(f){\le}1$; computable
$\mu$ have \trm {universal} (\ie majorant {\em r.e.}) tests $\T_\mu(\a)
{=}\sum_i\m(\a_{[i]})/\mu(\a_{[i]})$, called \trm {Martin-L\"of tests.}\footnote
 {The condition $\mu(\T_\mu){\le}1$, slightly stronger (in log scale)
  than the original one of \cite {ML}, was\\ required in \cite{L76}
  in order to satisfy conservation of randomness. Both types of tests
  diverge simultaneously.\\ \cite {Schn73} (for divergence of $\T_\l$),
  \cite {L73}, \cite {g80} characterized the tests in complexity terms.}
 Indeed, let $t$ be an \re $\mu$-test, and $S_k$ be an \re family of
prefixless subsets of $\St$ such that $\cup_{x\in S_k}x\W=\{\a:t
(\a){>}2^{k+1}\}$. Then $t(\a)=\Theta(\sum_{k,x{\in}S_k}(2^k[\a{\in}x\W]))
=\Theta(\sup_{k,x{\in}S_k}(2^k[\a{\in}x\W]))$. Now, $\sum_{k,x{\in}S_k}
(2^k\mu(x)) <\mu(t)\le1$, so $2^k\mu(x){=}O(\m(x))$ for $x{\in}S_k$
and $t(\a){=}O(\sup_{k,x{\in}S_k}([\a{\in}x\W]\m(x)/\mu(x))){=}
O(\sup_i(\m(\a_{[i]})/\mu(\a_{[i]})))$.

\trm{Martin-L\"of random} are $\a$ with finite \trm{rarity}
$\d_\mu(\a)\edf\|\ceil {\T_\mu(\a)}\|\eqa\sup_i(\|\mu(\a_{[i]})\|-\K(\a_{[i]}))$
and we also use $\d_\mu(\a|x)\edf\sup_i(\|\mu(\a_{[i]})\|-\K(\a_{[i]}|x))$.

\p {Continuous transformations} $A:\W{\to}\W$ induce normal linear
operators $A^*:f{\mapsto}g$ over $\Es$, where $g(\w){=}f(A(\w))$. So obtained,
$A^*$ are \trm {deterministic}: $A^*(\min\{f,f'\})=\min\{A^*(f),A^*(f')\}$.
Operators that are not, correspond to probabilistic transformations (their
inclusion is the benefit of the dual representation), and $g(\w)$ is then the
expected value of $f(A(\w))$. Such $A$ also induce $A^{**}$ transforming input
distributions $\mu$ to output distributions $\ph=A^{**}(\mu):\ph(f)=\mu(A^*(f))$.
 I treat $A,A^*,A^{**}$ as one function $A$ acting as $A^*$, or $A^{**}$ on the
respective (disjoint) domains. Same for partial transformations below and their
concave duals. I also identify $\w{\in}\W$ with measures $f\mapsto f(\w)$.

\section {Partial Operators, Semimeasures, Complexity of Prefixes}

Not all algorithms are total: narrowing down the output to a single
sequence may go slowly and fail (due to divergence or missing
information in the input), leaving a compact set of eligible results:

\BD\label{op}\BE \item 
{\em Partial} continuous transformations \trm {(PCT)} are compact subsets
$A\subset\W{\times}\W$ with $A(\a)\edf\{\b:(\a,\b){\in} A\}\ne\emptyset$.
When not confusing I identify singletons $\{\b\}$ with $\b{\in}\W$.\\
\trm{Computable} PCT are r.e., \ie enumerate the open complement of $A$;
 \item a PCT $A$ is \trm{clopen} if co-images $A^{-1}(s)=\{\a: A(\a)\subset s\}$
 of all clopen $s\subset\W$ are clopen.\\ $A$ is $t$-clopen
 if $A^{-1}(x\W)$ depend only on $\a_{[t(x_{[i]})]}$ for some $i$.
 \item\trm {Dual} of PCT $A$ is the operator $A^*:\Es\to\Ess$,
   where $A^*(f)=g:\a\mapsto\min_{\b{\in}A(\a)}f(\b)$. \EE\ED

An important example is a \trm {universal} algorithm $U$.
It enumerates all algorithms $A_i$ with a prefixless set $P$ of
indexes $i$ and sets $(i\a,\b)\in U$ iff $(\a,\b){\in}A_i,i{\in}P$.

\BR\label{cmp} Composing PCT with linear operators produces normal concave
operators, all of them by Hahn–Banach theorem. Indeed, each such $C(f)$ is a
composition $A(R(B(f)))$: Here a PCT $A(\a)$ relates each $\a$ to the binary
encodings $\{\mu\}$ of measures $\mu\ge C(\a)$; $R$ transforms $\{\mu\}$ into a
distribution $\{\mu\}\otimes\l$; and $B(\{\mu\}, \b)$ relates $\l$-distributed
$\b$ to $\mu$-distributed $\g$ with $\mu[0,\g)\le\b\le\mu[0, \g]$.\ER

Normal concave operators transform measures into \trm {semimeasures}:

\BD\label{sm}\BE\item A \trm {semimeasure} $\mu:\Es{\to}\R$ is
a normal ($\mu(\pm\one){=}{\pm}1,\,\mu(|f|){\ge}0$) functional\\
that is concave: $\mu(cf{+}g)\ge c\mu(f){+}\mu(g),\,c\in\R^+$,
\eg $\mu(x)\ge\mu(x0){+}\mu(x1)$, for $x\in\St$.\\ $\mu$ extends to
$f{\in}{-}\Ess$ as $\inf\{\mu(g):f\le g{\in}\Es\}$, and to other functions
as $\sup\{\mu(g): f\ge g{\in}{-}\Ess\}$, as is usual for inner measures.
$\mu$ is \trm {deterministic} if $\mu(\min \{f,g\})=\min\{\mu(f),\mu(g)\}$.

\item Normal ($A(\pm\one)=\pm\one$, $A(|f|)\ge0$) concave operators
$A:\Es\to\tld\Es$ transform input points $\a$ and distributions $\ph$ (measures
or semimeasures) into their output distributions $A(\ph):f{\mapsto}\ph(A(f))$.
Operators $A$ are deterministic if semimeasures $A(\a)$ are.\\
\trm {Regular} are semimeasures $A(\l)$ for deterministic \re $A$;
$t$-regular for a $t$-clopen $A$. \EE\ED

\BP\label{id}\BE\itemsep0pt\item\label{id1} Each deterministic $\mu$
is $\mu(f)=\min_{\w\in S}f(\w)$ for some compact $S\subset\W$. \item\label{id2}
Dual of PCT are those and only those operators that are normal, concave,
and deterministic.
\item\label{id3} Each $f{\in}\Es$ has a unique form $f{=}\sum r_if_i$
with distinct boolean $f_i{\ge}f_{i+1},f_0{=}\one$, $r_i{>}0$ for $i{>}0$.\\
Then $\un\mu(f)\edf\sum_i r_i\mu(f_i)$. $\mu{=}\un\mu$
 if $\mu$ is regular. All \re measures are regular.
 \item\label{id4} Each \re semimeasure $\mu$ has a regular
 \re $\mu'{\le}\mu$ with $\mu'(x){=}\mu(x)$ for all $x\in\St$.\\
$\mu'$ is $t$-regular for a computable $t$ if $\mu(x)$ have $<t(x)$ bits.\EE\EP

\BPR \ref{id1}: Note, $p(\b)\edf\inf_{g:\mu(g){\ge}1}|g(\b)|\in\{0,1\}$.
Indeed, if $\mu(f){-}f(\b)=t{>}0$ and $g=(f{-}f(\b)\one)/t$ then $g(\b){=}0$,
$\mu(g){\ge}1$. Then $S$ is $\{\b:p(\b){=}1\}$. \ref{id2}: $\mu{=}A(\a)$ are
deterministic, so $\mu(f){=}\min_{\b\in S}f(\b)$. \ref{id3} is since regular
$\mu$ are averages of deterministic ones. \ref{id4} is by Theorem 3.2 of
\cite{ZL}.\EPR

\subsection {Complexity: General Case}

\BP\label{um} There exists a \trm {universal}, \ie majorant (on $\Es^+$) \re
semimeasure {\M}. The values $\M(x)$ can have $\K(x)$ bits. (Thus $t$-clopen
PCT can generate $\M$ for any computable $t(x){>}\K(x)$).\footnote
{For $t(x)\sim\|x\|$ shown in \cite{L71}, Th.13;
also mentioned in \cite {ZL}, Prp.3.2.}\EP

\BPR For an \re family $\mu_i$ of all \re semimeasures, take $\M(x)=
\sum_i\mu_i(x)/2i^2$. $\M(x)$ can be rounded-up to $\K(x)$ bits after adding
$\sum_{y\ne\{\}}\m(xy)$ (to keep $\M(x)\ge\M(x0)+\M(x1)$).\EPR

 As in \cite {ZL}, $\KM(x)\edf\|\M(x\W)\|$.
 Same for $\M_\a$, \re w.r.t. $\a$ and $\KM(x|\a)\edf\|\M_\a(x\W)\|$.

$\K(x|y)$, $\KM(x)$ are examples of the many types of complexity measures on
$\St$.\\ \cite{L76b} gives the general construction of Kolmogorov-like
complexities $\K_v$. I summarize it here.

$\K_v$ are associated with classes $v$ of functions $m{:}\;\St{\to}[0,1]$, in
linear scale, and their logarithmic scale projections $\ov v\edf\{K=\|m\|:
m{\in}v\}$. Thus, $\K(x|y)$ is $\K_v$ for $v=\{m:\sup_y\sum_xm(x|y)\le1\}$.

These $v$ are closed-down, weakly compact, and decidable on tables with finite
support. $\ov v$ will have a minimal, up to $\eqa$, co-\re function $\K_v$.
This justifies the logarithmic scale where the values of $\K_v$ are well
defined up to $O(1)$ adjacent integers. (Though linear scale is often clearer
analytically.)

$\K_v$ minimality requires $\min\{K',K''\}{+} O(1)\in\ov v$ for any $K',K''$
in $\ov v$. In the linear scale of $m$ this comes
to $(m'{+}m'')/c\in v$ for some $c{=}O(1)$. I tightened this to convexity
with $c{=}2$; this changes $K$ in $\ov v$ by just $\Theta(1)$ factors:
a matter of choosing bits as units of complexity.

Similarly to Proposition~\ref{um}, this condition suffices for $\ov v$ to have
a minimal, up to $\eqa$, co-\re $\K_v$. Each such $\K_v{<}\infty$ has a
computable lower bound $B_v(x)=\min_{K{\in}\ov v}K(x)$, largest up to $\eqa$,
among \re bounds. And $\K_v{-}B_v$, too, is such a $\K_{v'}$; I call $v'$ \trm
{normal}, as $B_{v'}{=}0$. Let $\Es_1{=}\Es^+\cap\{f{:}\:\max_\a f(\a){=}1\}$.
$\KM(f){=}\|\M(f)\|,\,f{\in}\Es_1$ is a normal complexity measure and all
others are its special cases:

\BP For each normal $v$ a computable representation $t_x\in\Es_1$
for $x\in\St$ exists such that $\K_v(x)\eqa\KM(t_x)\lea\K(x)$.\EP

\BPR $\K_v(x){\lea}\K(x)$ follows from normality ($B_v{\eqa}0$) and convexity
of $v$. Thus $\m_v(x)$ needs $\lea\K(x)\lea2\|x\|$ bits. Let $m'$ be $m{\in}v$
so rounded-down. For $m{\in}v$, let $m_x$ be a prefixless code of $(x,m'(x))$,
and $m_{[x]}$ be $m_1m_2\ldots m_x$. Then $t_x(\a)\edf m'(x)$ if
$\a{=}m_{[x]}\b,m{\in}v$; otherwise $t_x(\a)\edf0$.

The measure concentrated in a single $\a$ has some $m{\in}v$ for which it maps
each $t_x$ to $m'(x)$.\\ Other measures $\mu$ also have $\tau_\mu:x\mapsto\mu
(t_x)$ in $v$ by convexity of $v$.\\ As $v$ is closed down, $\tau_\M\in v$, too,
and so, $\tau_\M=O(\m_v)$. Conversely, some measure $\a$ has $\tau_\a{=}\m_v$.
As $\m_v$ is r.e., the minimal semimeasure $\mu$ with $\tau_\mu\ge\m_v$ is
r.e., too, and so, $\m_v\le\tau_\mu=O(\tau_\M)$.\EPR

\section {Complete Sequences} \label{cmpl}

\cite{L76a} calls \trm {complete} sequences $\a$ that are $\mu$-random for a
computable $\mu$. This class is closed under all total recursive operators.
Here I use this term \trm {complete} also for $\a'$ Turing-equivalent to such
\a. This is identical to $\a'$ being either recursive or Turing-equivalent to a
$\l$-random sequence.

By \cite {ku,g86,BL}, each $\a{\in}\W$ is w.t.t.-reducible to a \l-random~\w.
Indeed, for $P(x,\a)=x\a$, let measure $\r$ be $\l$-integral of $\T_\l$:
$\r=P(\m\otimes\l)$. Let $R=\{\a:\T_\l(\a)\le c\}$ for a convenient constant
$c$. When $A(\l)$ generates $\M(x)$, the co-images of all prefixes intersect
$R$. (Otherwise $A(\r)$ would exceed $\M=A(\l)$.) But for clopen $A$ (see
Prp.~\ref{um}), co-image of any $\a{\in}\W$ is the intersection of (non-empty
in $R$) clopen co-images of its prefixes $\a_n$, so intersects $R$, too.

Yet partial algorithms can generate incomplete sequences with positive
probability: \cite {Vyugin}.

I extend $\K(\b|\a)$ to $\a,\b\in\W$ using a universal PCT $U(p,\a)$ that runs
on $\a$ a program $p$ given on a separate tape; $\a_p$ combines bits of $p,\a$
in order read by $U$. $p$ must be prefixless: $U$ diverges and $\a_p$ is
undefined unless $U$ detects the end of $p$ and does not try to move beyond
its end of tape.

\BD Here $\a,\b\in\W$. $\K(\b|\a)\,\edf\,\min_p\{\|p\|:U(p,\a){=}\b\}$.\\
The \trm {codeset} $R_\a$ for $\a$ is $\{\b:U(\b){=}\a,\,\d_\l(\b){<}c\}$
where $c$ is a constant such that\\ the \trm {incompleteness}\footnote
 {For some applications of $\Ks$ its lower bound $\|\M_\a(R_\a)\|$ may suffice.}
 $\Ks(\a)\edf\min_{\b\in R_\a}\K(\b|\a)$ of
 any $\a$ is $\lel\|\d_\l(\a)\|$\footnote
 {By finding $p$ to replace a prefix $q{=}U(p)$ where $\|q\|{-}\|p\|$
  is the rarity. \newcounter{cmpr}\setcounter{cmpr}\thefootnote}.\\
 \trm {Tight complexity} $\Ki(x|\a)$ is $\|\mf(x|\a)\|$ where $x{\in}\N$,
 $\widehat\m_x(\a)\edf\min_{\b\in R_\a}\m(x|\b)$,
 $\mf(x|\a)\edf\widehat\m_x{\uparrow}(\a)$.\ED

 These concepts satisfy many properties similar to
 those given (for integers) in \cite {g74,L74}:

 \BP\label{kfs}\BE\itemsep0pt \item\label{kf1} $\K(\b|\a)\sim\KM(\b|\a)$.
 \item\label{kf2} $\d_\l(\b_q)\eqa\d_\l(\b)+\|q\|{-}\K(q|\b,\d_\l(\b))$.
 \item\label{kf3} $\Ks(\a)\eqa\min_{\b}\{\K(\a|\b){+}\K(\b|\a){+}\d_\l(\b)\}$.
\item\label{kf4} $\Ki(x|\a)\eqa\Ki(\ov x|\a)$. (Recall: $\ov x$ is $(x,\K(x))$.)
 \item\label{kf5} $\If(\a:x)\edf\K(x)-\Ki(x|\a)\lea\If(\a:(x,y))$.
 \item\label{kf6} $\If(\a:x)\eqa(\min_{\b\in R_\a}\d_\l(\b|\ov x)){\uparrow}
        \eqa(\min_{\b\in U^{-1}(\a)}\d_\l(\b|\ov x)){\uparrow}$.\EE\EP

 \BPR\ref{kf1}.
Let $k{=}\KM(\b|\a)$, $s_{k,\a}\edf\{x0,x1:\KM(x0\W|\a){<}k,\KM(x1\W|\a){<}k\}$,
 so, $|s_{k,\a}|<2^k$.\\ Let $x$ be the longest prefix of $\b$ in $s_{k,\a}$.
 Then $\K(x|\a,k)\lea k$, and $\b$ can be computed from $x,k,\a$.

 \ref{kf2}. "$\d_\l(\b_q)\gea$'' is by $t_{\b_q}\edf\T_\l(\b)2^{\|q\|}
 \m(q|\b,\d_\l(\b))$ being \re with $\l_{(\b_q)}(t_{\b_q})\le1$. For "$\lea$''
 take a distribution $\mu_{\b,d}(q)\edf\T_\l(\b_q)/2^{\|q\|+d}$ enumerated
 for each $\b,d$ only while $\dl_\b\edf\|\sum_q 2^d\mu_{\b,d}(q)\|{\le}d$;\\
 so enumeration of $\mu_{\b,\dl_\b}$ is not stopped.
 Now, $\dl_\b\eqa\d_\l(\b)$ since $\l_{(\b)}(2^{\dl_\b}){\le}1$.
 Also, $\sum_q\mu_{\b,d}(q){=}O(1)$, so $\mu_{\b,d}(q){=}O(\m(q|\b,d))$. Thus,
 $\d_\l(\b)+\|q\|-\d_\l(\b_q)\gea\|\mu_{\b,\dl_\b}(q)\|\gea\K(q|\b,\d_\l(\b))$.

 \ref{kf3}. Take $p,q,\b{=}U(p,\a)$ with $U(q,\b){=}\a$, $\Ks(\a){\eqa}
 \|p\|{+}\|q\|{+}\d_\l(\b)$.\\ Then $\d_\l(\b)\eqa0$, $\K(q|\b)\eqa\|q\|$,
 else $\b$ or $q$ could be shrunk decreasing $\Ks(\a)$.\\
 Then $\d_\l(\b_q)\eqa0$ by \ref{kf2}, and the claim follows
 by appending $q$ to $p$ to map $\a\mapsto(q,\b){\mapsto\b_q}$.

 \ref{kf4}. Let $\b{=}v\w,\,\d_\l(\b){\eqa}0,\,\|p\|{=}\K(x|\b)$
 (and so, ${\eqa}\K(p|\b)$), and $U(p,\b){=}x$ reads only $p,v$, so,\\
 $\K(p,v){\lea}\|pv\|$. Then $\|pv\|{-}\K(p,v)\lea\d_\l(v_p){\eqa}0$
 by \ref{kf2}. So, $\K(x){+} \K((p,v)|\ov x)\eqa\K(p,v)\eqa\|pv\|$.\\
 Thus, finding $i,j$ with $\K(x){<}i,\K((p,v)|x,i){<}j$,
 $i{+}j{\lea}\|pv\|$ computes $\K(x)\eqa i$ from $p,v$.

 \ref{kf5}. By \ref{kf4} and $\K(\ov x|\ov{(x,y)}){\eqa}0$, we can replace $x$
with $\ov x$. Let $\d_\l(\b){\eqa}0$.\\ Then $\K(\ov x)-\K(\ov x|\b)-\K(\ov x,y)
+\K((\ov x,y)|\b)\eqa\K(y|\b,\ov x,\K(\ov x|\b)){-}\K(y|\ov x)\lea0$.

 \ref{kf6}. For $\b{\in}R_\a$, $\K(x)-\K(x|\b)\gea\d_\l(\b|x)$, \ie
 $\m(x)\T_\l(\b|x)=O(\m(x|\b))$.\\ Indeed, the \re $\sum_x\m(x)\T_\l(\b|x)$
 is $O(\T_\l(\b)){=}O(1)$ since $\l_{(\b)}(\sum_x\m(x)\T_\l(\b|x))=\\
 \sum_x\m(x)\l_{(\b)}(\T_\l(\b|x))\le\sum_x\m(x){\le}1$.
 Also for all $\b$, $x{=}U(p)$ with $\K(x){=}\|p\|{\eqa}\K(p)$, the \re\\
 $\m(U(p)|\b)p$ is $O(\T_\l(\b|p))$ since $\l_{(\b)}\m(x|\b)/\m(x){=}O(1)$.
 So and $\K(x)-\K(x|\b)\lea\d_\l(\b|\ov x)$.\\
 And any $\b\in U^{-1}(\a)$ can be compressed\footnotemark[\thecmpr]
 to $\b'{\in}R_\a$ with $\d_\l(\b'|\ov x)\lea\d_\l(\b|\ov x)$.\EPR

\vspace{1pc}\section {Rarity}

 \subsection {Non-algorithmic Distributions}

\cite {L73} considered a definition of rarity $\T_\mu(\a)$ for arbitrary
measures $\mu$ where $\T_\mu$ is \re only relative to $\mu$ used as an oracle.
This concept gives interesting results on testing for co-\re classes of
measures such as, \eg Bernoulli measures. Yet, for individual $\mu$ it is
peculiar in its strong dependence on insignificant digits of $\mu$ that have
little effect on probabilities. \cite {L76,g80} confronted this aspect by
restrictions making $1/\T_\mu(\a)$ monotone, homogeneous, and concave in
$\mu$.\footnote
 {The Definition in \cite {L76} has a typo: "$Q(f)$'' meant to be "$Q(g)$''.
 Also, in English version "concave relative to $P$''
 would be clearer as "for any measure $Q$ concave over $P$''.
 So, its $\T_\mu(\a)$ is $\sup_{f,g\in\Es}(t(f|g)f(\a)/\mu(g))$,
 for\\ a $t$ majorant among \re functions that keep $\T_\mu(\mu)\le1$
 for all measures $\mu$, where $\T_\mu(\ph)\edf\ph_{(\a)}(\T_\mu(\a))$.\par
 Restrictions on $t$ (\eg $t\subset\St{\times}\Es$, $\T_\mu(\a)\edf\sup_
{(f,g){\in}t}f(\a)/\mu(g)$) can reduce redundancy with no loss of generality.}

\cite {L84} used another construction for $\T_\mu(\a)$. It generates
$\mu$-tests by randomized algorithms and averages their values on $\a$. For
computable $\mu$ the tests' ${\le}1$-mean can be forced by the generating
algorithm, so the definition agrees with the standard one. But for other $\mu$
the ${\le}1$-mean needs to be imposed externally. \cite {L84} does this by just
replacing the tests of higher mean with $\one$ (thus tarnishing the purity of
the algorithmic generation aspect). That definition respects the conservation
inequalities, so for \re semimeasures it gives a lower bound for our
$\d_\mu(\a)$ below (by Prop.\ref{mx}).

\subsection{R.E. Semimeasures}

\p {Coarse Graining.} I use $\l$ as a typical continuous computable
measure on \W, though any of them can be equivalently used instead. Also, any
recursive tree of clopen subsets can serve in place of $\St$.

Restricting inputs $\w$ of a PCT $A$ to those with converging outputs (\ie a
singletons $A(\w)\in\W$) truncate the output semimeasure to a smaller {\em
  linear} functional: a maximal measure $\mu^\Es\le\mu{=}A(\l)$. Yet, much
information is lost this way: \eg $\|\M^\Es(x)\|,x{\in}\St$ has no recursive in
$\|\M(x)\|$ upper bound. To keep information about generated prefixes, I will
require linearity of $\mu^E$ only on a subspace $E{\subset}\Es$. $E$ will play
a role of space of $\mu^E$-measurable functions. E.g., relaxing $A(\w)$
restriction from singletons to sets of radius ${\le}2^{-n}$, produces a
semimeasure linear on the subspace of $f$ with $f(\a)$ dependent only on
$\a_{[n]}$. Subspaces $E\subset\Es$ used below are generated by
subtrees\footnote
 {If a non-binary tree is used instead of $\St$ then any
  $x{\in}S$ must have either all its children in $S$ or none.}
 $S{\subset}\St$, \ie are spaces of linear combinations of functions in $S$.
 By \trm {$E$-measures} I call semimeasures linear on such $E$.

\BP\label{cg} Each semimeasure $\mu$, for each $E$,
 has the largest (on $\Es^+$) $E$-measure $\mu^E\le\mu$.\EP

\BPR Let $X$ be the set of all measures $\ph$ which, for some
$F{\subset}E^+$ with $\sum_{f\in F}f>0$\\ and all $g\in\Es^+$, $g\le f\in F$,
have $\ph(g)\ge\mu(g)$. Then $\mu^E(f)=\inf_{\ph\in X}\ph(f)$.\EPR

Now, I will extend the concept of rarity $\T_\mu$, $\d{\edf}\|\ceil{\T}\|$
from computable measures $\mu$ to r.e. semimeasures. The idea is for
$\d_\mu(\a)$ to be bounded by $\d_\l(\w)$ if $\a{=}A(\w)$, $\mu{\ge}A(\l)$.
Coarse graining on a space rougher than the whole $\Es$, allows to define
rarity not only for $\a{\in}\W$ but also for its prefixes. For semimeasures,
rarity of extensions does not determine the rarity of a prefix.

$\T_\mu$ for a computable measure $\mu$ is a single \re function $\W\to\R^+$
with $\le1$ mean. It is obtained by averaging the \re family of all such
functions. This fails if $\mu$ is a semimeasure: its mean of sum can exceed the
sum of means. So, our extended $\T_\mu$ will be refined with a subspace
$E{\subset}\Es$ parameter.

\BD\label{d1} For an $E{\subset}\Es$ and a PCT $A$,
$t^E_A$ is $\sup\{f{\in}E:A(f)\le\T_\l\}$.\ED

\BP\label{uo} Each \re $\mu$, among all \re PCT $A$ with $A(\l)\le\mu$,
has a universal one $U_\mu$, \ie such that $t^E_{U_\mu}=O(t^E_A)$ for each $A$
and all $E$. $\mu(f)\le\l(2U_\mu(f))$ if $f\in\St$ or $\mu$ is regular.\EP

\BPR $U(i\w)\edf A_i(\w)$ for a prefixless enumeration $A_i$ of all such $A$.
\EPR

\BD\label{rm} $\T_\mu^E(\ph)$ for semimeasures $\ph$, \re $\mu$ is the
mean: $\ph^E(t^E_{2U_\mu})$ for $U_\mu$ defined in Prop.\ref{uo}.\ED

\BL\label{dM} (1) $\d_\mu^\Es\eqa\d_\mu$ for computable measures $\mu$.
 (So, if $E=\Es$, we omit $E$ in $\d_\mu^E\edf\|\ceil{\T_\mu^E}\|$.)\\
 (2) $\d_\mu^E(\mu){=}0$.\hspace{3pc}
 (3) $\d_\M\eqa0$ for the universal semimeasure $\M$.\EL

\BPR (1) follows from \cite {ZL} Th.~3.1 and enumerability of $\T_\mu$.

(2) Let $A{=}U_\mu$. By Prop.\ref{uo}, $\mu^E(f)/2\le\l(A(f))$
for $f{\in}\St$, and thus for $f{\in}E^+$. Also any $f{<}t_A^E$ is
$<\sum_if_i$ where $f_i{\in}E^+$, $f_if_{j{\ne}i}=0$, and $A(f_i)\le\T_\l$.
Now, $\T_\mu^E(\mu)=\sup_{f{\in}\Es^+,f{<}t_A^E}\mu^E(f)/2$,\\ and $\mu^E(f)/2
\le\sum_i\mu^E(f_i)/2\le\l(\sum_i A(f_i))=\l(\sup_i A(f_i))\le\l(\T_\l)\le1$.

(3) By \cite{g86,ku}, an \re PCT $A$ exists such that any {\a} is $A(\w)$ with
$d_\l(\w){=}0$. Then $g{=}A(f)\le\T_\l$ means $g(\w){=}f(A(\w))=f(\a)\le\T_\l(\w)
\le2$. For a universal $\M$, $\d_\M\lea\d_{A(\l)}\eqa0$.\EPR

Let the semimeasure $\nu{=}\mu{\otimes}\ph$ on $\W^2$ be the minimum of
$\mu'{\otimes}\ph'$ over all measures $\mu'{\ge}\mu,\,\ph'{\ge}\ph$.
Then $\nu(h){=}\mu(f)\ph(g)$ for $h(\a,\b){=}f(\a)g(\b)$, and for all $h$, if
$\ph$ is a measure, $\nu(h){=}$ $\mu(\ph_{(\b)}(h(\a,\b)))$. Let $E\otimes\Es$
be the space generated by $\{f(\a)g(\b),\,g{\in}E,f{\in}\Es\}$.
Adding coin-flips preserves randomness:

\BL\label{cir} $\d_{\mu{\otimes}\l}^{E\otimes\Es}(\ph{\otimes}\l)\lea
\d_\mu^E(\ph)$ for all $\ph$, \re $\mu$, space $E{\subset}\Es$.\EL

\BPR Let $\phi\edf\ph{\otimes}\l$, $\nu\edf\mu{\otimes}\l$, $E'\edf
E{\otimes}\Es$, $A(\a,\b)\edf(U_\mu(\a),\b)$, $t\edf\T_\nu^{E'}(\phi)=
\phi^{E'}(t^{E'}_{U_\nu})$. Then\\ for some $c{\in}\Q^+$,
$t/c<\phi^{E'}(t^{E'}_{A})=\phi^{E'}(\sup H)$ where $H=\{h{\in}E':A(h)
{\le}\T_{\l^2}\}$. So $t/c<\phi^{E'} (\sup G)$ for a finite set
$G=\{f_i(\a)g_i(\b)\}\subset H$ with $\l(g_i){=}1$ and $f_if_{j\ne i}{=}0$,
thus $\sup G=\sum G$.\\ Now, $U_\mu(f_i)g_i<\T_{\l^2}$,
 thus $U_\mu(f_i)<\l_{(\b)}(\T_{\l^2}(\a,\b))=O(\T_\l(\a))$. 
 Then, $t/c<\phi^{E'}(\sum_if_i g_i)=\sum_i\phi^{E'}(f_i g_i)=\sum_i\ph^E(f_i)=
\ph^E(\sum_if_i)=\ph^E(\sup_if_i)=O(\ph^E(t^E_{U_\mu}))=O(\T_\mu^E(\ph))$.\EPR

Let $A(E)$ be $\{f{\in}\Es:A(f){\in}\tld E{\subset}\Ess\}$.
Deterministic processing preserves randomness, too:

\BL\label{cid} $\d_{A(\mu)}^{A(E)}(A(\ph))\lea\d_\mu^E(\ph)$ for
 each \re PCT $A$, all $\ph$, \re $\mu$, space $E{\subset}\Es$.\EL

 \BPR Let $E'\edf A(E)$, $\phi\edf A(\ph)^{E'}\le A(\ph^E)$,
 $A_\mu(f)\edf U_\mu(A(f))$. So, $t\edf\T_{A(\mu)}^{E'}(A(\ph))=\\
 \phi(t^{E'}_{U_{A(\mu)}})<c\,\phi(t^{E'}_{A_\mu})<c\,\phi(\sup F)$
 for $F\edf\{f{\in}E'^+:U_\mu(A(f))\le\T_\l\}$ and some $c\in\Q^+$.\\
 Then $t<c\phi (\sup G)$ for a finite set $G\subset F$ that can
 be made disjoint, \ie $gg'=0$\\ for $g{\ne}g'$ in $G$ (and thus
 $A(g)A(g')=0$ as $A$ is deterministic), so $\sup G=\sum G$.\\
 Now, $U_\mu(h){\le}\T_\l$ for $h\edf\sup\{A(f):f{\in}F\}{\in}\tld E^+$,
 so $h\le t^E_{U_\mu}$. Then $t/c<\phi(\sup G)= \phi(\sum G)=\\
 \sum_{g{\in G}}\phi(g)\le \sum_{g{\in G}}\ph^E(A(g))= \ph^E(\sum_{g{\in G}}A(g))=
 \ph^E(\sup_{g{\in G}}A(g)) \le \ph^E(h) \le 2\T_\mu^E(\ph)$. \EPR

By the remark~\ref{cmp}, Lemmas~\ref{cir}, \ref{cid} imply the following
theorem:

\BT[Randomness Conservation]\label{t1} The test $\d$ satisfies
 $\d_{A(\mu)}^{A(E)}(A(\ph))\lea \d_\mu^E(\ph)$\\ for each normal concave
 \re operator $A$, all $\ph$, \re $\mu$, space $E{\subset}\Es$.\ET

 These tests $\d_\mu^E$ are the strongest (largest)
 extensions of Martin-L\"of tests for computable $\mu$:

\BP $\T_\mu^E(\w)$ is majorant among extensions $\tau_\mu {\in}\tld E^+$
of Martin-L\"of test $\T_\l=\tau_\l$\\ that are non-increasing
on $\mu$ and obey Lemma~\ref{cid} for $\|\ceil\tau\|$
with $\tau^E_\mu(\ph)\edf\ph^E(\tau_\mu)$.\label{mx}\EP

\BPR With $A{\edf}U^*_\mu$, $A(\tau_\mu){\le}A(\tau_{A(\l)})$ and Lemma~\ref
{cid} for $\|\ceil\tau\|$ gives $A(\tau_{A(\l)})(\w)=\tau_{A(\l)}(A^*(\w))\le
c\,\tau_\l(\w){=}c\T_\l(\w)$ for some $c\,{\in}\Q^+$. If $\tau_\mu{>}2c\,f{\in}
E^+$ then $2c\,A(f){<}A(\tau_\mu){\le}c\T_\l$, so $\T_\mu^E{>}f$ as defined.\EPR

\section {Information and its Bounds}

Now, like for the integer case, mutual information $\I(\a:\b)$ can be
defined as the deficiency of independence, \ie rarity for the distribution
where $\a,\b$ are assumed each universally distributed (a vacuous
assumption, see \eg Lemma~\ref{dM}(3)) but independent of each other:
\[\I(\a:\b)\edf\d_{\M\otimes\M}((\a,\b)).\]
 Its conservation inequalities are just special cases of Theorem~\ref{t1}
and supply $\I(\a:\b)$ with lower bounds $\I(A(\a):B(\b))$
for various operators $A,B$. In particular transforming $\a,\b$ into
distributions $\m(\cdot|\a),\m(\cdot|\b)$, gives $\I(\a:\b)\gea\i(\a:\b)\edf
\|\ceil{\sum_{x,y\in\N}\m(x|\a)\m(y|\b)2^{\I(x:y)}}\|$.\footnote
 {This $\i$ was used as the definition of information in \cite {L74}.}
 Same for $\If(\a:\b)\edf\|\ceil{\sum_{z\in\N}
 \mf(z|\a)\mf(z|\b)/\m(z)}\|\gea\i(\a:\b)$.\footnote
 {$\If\gea\i$ since for $z{=}(x,y)$, by Prop.\ref{kfs}.\ref{kf4},
 $\Ki(z|\a)\lea\Ki(\ov y|\a){+}\K(x|\ov y)\eqa
 \Ki(y|\a){+}\K(x|\ov y)\lea\K(y|\a){+}\K(x,y){-}\K(y)$.} These bounds also
satisfy the conservation inequalities, and agree with $\I(\a:\b)$ for $\a,
\b\in\N$. While $\I$ is the largest such extension from $\N$, $\i$ is the
smallest one. Interestingly, not only for integers, but also for all complete
sequences this simple bound $\i$ is tight, as is an even simpler one
$\i'(\a:\b)\edf\sup_{x\in\N}(\K(x){-}\K(x|\a){-}\K(x|\b))\lea\i(\a{:}\b)$:

\BP\label{cmpl-i} For $\a,\b\,{\in}\W,\,b\,{\in}\N$:
 (1) $\I(\a:b)\eqa\K(b){-}\Ki(b|\a)$ (follows from Prop.\ref{kfs}.\ref{kf6});\\
 (2) $\I(\a:\b)\lea(\min_{\a'\in R_\a,\b'\in R_\b}\i'(\a':\b')){\uparrow}
      \lea\i'(\a:\b)+\Ks(\a){+}\Ks(\b)$.\EP

In particular, this can be used for $\a$ being the Halting Problem sequence
(which is complete, being Turing-equivalent to any random \re real,
such as, \eg one constructed in sec.~4.4 of \cite {ZL}).

\BPR We can replace $\a,\b$ with $\a'{\in}R_\a,\b'{\in}R_\b$.
 Let~$h_n\edf(\a_{[n]},\b_{[n]})$.\\ $\l^2\edf\l{\otimes}\l\,{=}\,O(\M^2)$,
so $\I(\a{:}\b){\lea}\d_{\l^2}((\a,\b)){\eqa}\|\ceil{\sup_n4^n\m(h_n)}\|
\eqa\sup_n(\K(h_n)-2(\K(h_n){-}n))$.\\ Also $t\edf\sum_{n,v}2^n\m((\a_n,v))=
\Theta(\T_\l(\a))$, so $2^n\m((\a_n,v))/t=O(\m((n,v)|\a,\|t\|))$, and\\
$\K(h_n|\a)-(\K(h_n)-n)\lel\|t\|\eqa0$. Thus $\K(h_n|\a)\lea\K(h_n)-n$ and
$\K(h_n|\b)\lea\K(h_n)-n$.\\ Then $\I(\a:\b)\lea\sup_n(\K(h_n)-2(\K(h_n){-}n))
\lea\sup_n(\K(h_n)-\K(h_n|\a){-}\K(h_n|\b))\lea\i'(\a:\b)$.\EPR

\BP\label{M} Let $A\subset\W$. Then $\M^\Es(A)=0$ iff
 $\exists\a\forall\b_{\in A}\I(\b:\a)=\infty$.\EP

\BPR "If'' is by Theorem\ref{t1}. Now, any $A$ with $\M^\Es(A){=}0$ has a
 sequence $\a$ of clopen sets $\a_i\subset\W$ with shrinking $\M(\a_i)$, \ie
 $\l(\{\g:\exists x\, U(\g)\subset x\W{\subset}\a_i\})<2^{-i}$, and s.t. each
 $\b{\in}A$ is in infinitely many $\a_i$. Then, by Prop.\ref{kfs}.\ref{kf6},
 $\If(\b:(i,\a_i))\gea(\min_{\g\in U^{-1}(\b)}\d_\l(\g|i,\a_i)){\uparrow}\gea i$
 and so $\I(\b:\a)\gea\If(\b:\a)=\infty$.\EPR


\newpage\begin{thebibliography}{99}\itemsep0pt\parskip2pt
 \bibitem[DAN]{DAN} {\em Doklady} AN SSSR = Soviet {Math.} Doclady.
 \bibitem[Barmpalias, Lewis-Pye 18]{BL} George Barmpalias, Andrew Lewis-Pye.
 2018.\\ Optimal redundancy in computations from random oracles. {\em
 J.Comp.Sys.Sci., \bf 92}:1-8.\\ Also: \hreff {arxiv.org/abs/1606.07910}
 \bibitem[Chaitin 75]{Chtn75}Gregory J. Chaitin. 1975.\\A Theory of Program-Size
 Formally Identical to Information Theory. {\em JACM}, {\bf 22}:329-340.
 \bibitem[G\'acs 74]{g74} Peter G\'acs. 1974.
 On the Symmetry of Algorithmic Information. \cite {DAN}, {\bf 15}:1477.
 \bibitem[G\'acs 80]{g80} Peter G\'acs. 1980.
 Exact expressions for some randomness tests.\\
 {\em Zeitschrift f. Math. Logik und Grundlagen d. Math.}, {\bf 26}:385--394.
 \bibitem[G\'acs 86]{g86} Peter G\'acs. 1986. Every Sequence is Reducible to a
 Random One.\\ {\em Inf.{\rm\&}Cntr.}, {\bf 70}/2-3:186-192.
 \bibitem[Kolmogorov 65]{K65} Andrei N. Kolmogorov. 1965.
 Three Approaches to the Concept\\ of the Amount of Information.
 {\em Probl.Pered.Inf.= Probl.Inf.Transm.}, {\bf 1}/1:1-7.
 \bibitem [Kucera 85]{ku} Antonin Kucera. 1985.
 Measure, $\Pi^0_1$-classes and complete extensions of PA.\\
 {\em Lecture Notes in Math., \bf 1141}:245–259. Springer.
 \bibitem[L 71]{L71} Leonid A. Levin. 1971. Some Theorems on the Algorithmic
 Approach to Probability Theory and Information Theory. Moscow University
 dissertation (in Russian).\\
 \hreff {www.cs.bu.edu/fac/lnd/dvi/diss/1-dis.pdf}\\
 English translation: {\em APAL}, {\bf 162}/3:224-235.
 \hreff {arxiv.org/pdf/1009.5894.pdf}
 \bibitem[L 73]{L73} Leonid A. Levin. 1973. On the Concept of a Random
 Sequence. \cite {DAN}, {\bf 14}/5:1413-1416.
 \bibitem[L 74]{L74} Leonid A. Levin. 1974. Laws of Information Conservation
 (Non-growth) and Aspects of the\\ Foundations of Probability Theory.
 {\em Probl.Pered.Inf.= Probl.Inf.Transm.}, {\bf 10}/3:206-210.
 \bibitem[L 76]{L76} Leonid A. Levin. 1976. Uniform Tests of Randomness.
 \cite {DAN}, {\bf 17}/2:337-339. 
 \bibitem[L 76a]{L76a} Leonid A. Levin. 1976. On the Principle of Conservation
 of Information\\ in Intuitionistic Mathematics. \cite{DAN}, {\bf 17}/2:601-605.
 \bibitem[L 76b]{L76b} Leonid A. Levin. 1976. Various Measures of Complexity for
 Finite Objects\\ (Axiomatic Description). \cite {DAN}, {\bf 17}/2:522-526.
 \bibitem[L 84]{L84} Leonid A. Levin. 1984. Randomness Conservation
 Inequalities. {\em Inf.{\rm\&}Cntr.}, {\bf 61}/1:15-37.
 \bibitem[L 13]{fi} Leonid A. Levin. 2013. Forbidden Information.
 {\em JACM}, {\bf 60/2}. \hreff {arxiv.org/abs/cs/0203029}
 \bibitem[L16]{L16} Leonid A Levin. 2016. Occam Bound on Lowest Complexity of
 Elements. \\{\em APAL}, {\bf 167}/10:897-900. \hreff {arxiv.org/pdf/1403.4539}
 \bibitem[Li, Vit\'anyi 08]{LV08} Ming Li, Paul Vit\'anyi. 2008.\\ {\em An
 Introduction to Kolmogorov Complexity and Its Applications.} Springer.
 \bibitem[Martin-L\"of 66]{ML} Per Martin-L\"of. 1966. On the Definition of
 Infinite Random Sequences.\\ {\em Inf.{\rm\&}Cntr.}, {\bf 9}:602-619.
 \bibitem[Schnorr 73]{Schn73} C.P. Schnorr. 1973. Process Complexity and
 Effective Random Tests.\\ {\em J.Comp.Sys.Sci.}, {\bf 7}:376-378.
 Also personal communication by Schnorr cited in \cite{Chtn75}.
 \bibitem[Solomonoff 64]{Sol} R.J. Solomonoff. 1964.
 A Formal Theory of Inductive Inference. {\em Inf.{\rm\&}Cntr.}, {\bf 7}/1.
 \bibitem[V'yugin 82]{Vyugin} Vladimir V. V'yugin. 1982.
 The Algebra of Invariant Properties of Binary Sequences.
 {\em Probl.Pered.Inf.= Probl.Inf.Transm.}, {\bf 18/2}, 147-161.
 \bibitem[ZL 70]{ZL} Alexander Zvonkin, Leonid A. Levin. 1970. The complexity
 of finite objects and the algorithmic concepts of information and
 randomness. {\em UMN = Russian Math.~Surveys}, {\bf 25}/6:83-124.
 \end{thebibliography}
 \end{document}